\documentclass[pre,twocolumn,aps,showpacs,supergroupedaddress,epsfig,amsmath,amssymb,eqsecnum]{revtex4}
\usepackage{epsfig,amsmath,amssymb,bm,epsf,graphics,graphicx,psfrag,verbatim,subfigure,framed}
\usepackage[usenames,dvipsnames]{color}
\usepackage{bbm}
\usepackage{mathrsfs}
\usepackage{amsfonts}
\usepackage{color}
\usepackage{pbox}
\def\newblock{\hskip .11em plus .33em minus .07em}

\def\debye{\ell_{{\rm DB}}}

\newcommand{\be}{\begin{equation}}
\newcommand{\ee}{\end{equation}}
\newcommand{\ba}{\begin{eqnarray}}
\newcommand{\ea}{\end{eqnarray}}
\newcommand{\bw}{\begin{widetext}}
\newcommand{\ew}{\end{widetext}}
\newcommand{\workone}{{\em paper I}}
\newcommand{\worktwo}{{\em paper I\!I}}

\newcommand{\rv}{{\bm{r}}}

\begin{document}
\title{Charged Plate in Asymmetric Electrolytes: One-loop Renormalization of Surface Charge Density and Debye Length due to Ionic Correlations}
\author{Mingnan Ding}
\author{Bing S. Lu}
\author{Xiangjun Xing}
\email{xxing@sjtu.edu.cn}
\affiliation{Department of Physics and Astronomy, and Institute of Natural Sciences, Shanghai Jiao Tong University, Shanghai, China}
\date{\today}
\begin{abstract}
The self-consistent field theory (SCFT) is used to study the mean potential near a charged plate inside a $m:-n$ electrolyte.  A perturbation series is developed in terms of $g = 4 \pi b/\ell_{\rm {\scriptscriptstyle DB}}$, where $b, \ell_{\rm{\scriptscriptstyle  DB}}$ are Bjerrum length and {\em bare} Debye length respectively.  To the zeroth order, we obtain nonlinear Poisson-Boltzmann theory.  For asymmetric electrolytes ($m \neq n$), the first order (one-loop) correction to mean potential contains a {\em secular term}, which indicates the breakdown of regular perturbation method.  Using a renormalizaton group transformation (RG), we remove the secular term and obtain a globally well-behaved one-loop approximation with {\em a renormalized Debye length} and {\em a renormalized surface charge density}.  Furthermore, we find that if the counter-ions are multivalent, the surface charge density is renormalized substantially {\em downwards}, and may undergo a change of sign, if the bare surface charge density is sufficiently large.
\end{abstract}

\maketitle
\section{Introduction}
\label{sec:Intro}
There is a general consensus \cite{Hansen:2000wj,Belloni:1998ez,shklovskii3,Levin-charge-review} that the Poisson-Boltzmann theory is inadequate in describing the statistical physics of electrolytes in the following situations: 1) near strongly charged surfaces; 2) in dense electrolytes; and 3) in asymmetric electrolytes.  The physical mechanisms have two candidates: 1) correlation effects, which are beyond PB manifestly, and 2) ion-specific interactions \cite{Kunz-specific-ion-effects-colloids, Lyklema:2006fk}, which are beyond the primitive model.  Because of the complexity of electrolyte systems, there does not yet exist a single theoretical framework capable of describing both non-PB aspects of electrolyte physics.

The present work is the third of a sequel that analyze the statistical physics of electric double layers (EDL) with planar geometry inside asymmetric electrolytes.  In Ref.~\cite{xing-ming} (which shall be referred to as {\em paper} I), M. Han and X. Xing solved the nonlinear Poisson-Boltzmann equation for a single strongly (and positively) charged plate inside a generic $m:-n$ electrolyte.  Using the leading order far field asymptotics of the mean potential, one can define a renormalized (or effective) charge density for the strongly charged plate, which saturates to a finite value that depending on valences $m,n$, as well as the ion density.  Note that this renormalization of surface charge density arises due to the nonlinearity inherent in the PB equation, which is a {\it mean field theory}.  There are also additional renormalization of charge density due to statistical fluctuations, which is completely ignored in the Poisson-Boltzmann theory.  Subsequently, in Ref.~\cite{xing-bing} (which shall be referred to as \worktwo ), two of us (B.S. Lu and X. Xing) calculated the correlation energy of a test ion near a strongly charged plate inside a $m:-n$ electrolyte, to the first order in $g$.  It was found that  for $m \neq n$, the correlation energy decays in the same fashion as the mean field potential in the far field.  This correlation energy was used to calculate the first order correction to the mean potential, which was found to contain a {\em secular} term that dominates the zeroth order result in the far field, indicating the breakdown of regular perturbation method.  In the present work, we shall use perturbation analyses and renormalization group method to demonstrate that the physical origin of the secular term is the renormalization of Debye length due to electrostatic correlations.  Additionally, we shall also obtain the renormalization of surface charge density due to the same fluctuations.

The remaining of this paper is organized as follows.  In Sec.~\ref{sec:formalism} we discuss the framework of self consistent field theory and perturbation method.  In Sec.~\ref{sec:symmetric}, we apply the method to the case of a charged plate inside symmetric electrolytes, and obtain the first order renormalized surface charge density. In Sec.~\ref{sec:asymmetric}, we study the (much harder) case of asymmetric electrolytes, and obtain the renormalized Debye length and renormalized surface charge density. To remove the secular terms, a technically challenging renormalization group analysis has to be carried out.  Finally in Sec.~\ref{sec:conclusion}, we summarize our work and discuss the implications of our results.  In Appendix \ref{2:-1&1:-2}, we present some analytic details about $2:-1$ and $1:-2$ asymmetric  electrolytes.
Our results are summarized by two equations (\ref{m:n alpha-2}) and (\ref{eta_R_mn}).

\vspace{-3mm}
\section{Formalism}
\label{sec:formalism}
\subsection{Self-consistent Field Theory (SCFT)}
As in \workone  and \worktwo, we shall consider asymmetric electrolyte with point-like positive/negatives ions carrying charges $+mq$ and $-nq$ respectively \footnote{Strictly speaking, two-component electrolytes with point like ions are not stable because opposite ions can approach infinitely close to each other so that the energy does not have a lower bound.  This pathology does not concern us because the divergence does not show up in our approximation. }.  The mean potential $\Phi(\rv)$ satisfies the exact Poisson equation:
\be
\label{eq:Poisson}
- \epsilon \, \nabla^2 \Phi(\rv) =
m q\, \rho^0_+ \, e^{- \beta w_1(\rv,mq)}
- n q\, \rho^0_- \, e^{- \beta w_1(\rv,-nq)},
\ee
where $\rho_{\pm}^0$ are the average ion number densities in the bulk, whereas $ w_1(\rv,mq),  w_1(\rv, -nq)$ are their potentials of mean force (PMF).
As discussed in the first section of \worktwo, $ w_1(\rv,kq)$ of a $k$-valence test ion can be formally expanded in terms of $k$:
\be
w_1(\rv, k q) = k q \, \Phi(\rv)
+ \frac{1}{2}  k^2 q \, \delta \Upsilon(\rv, \rv)
+ O(k^3),
\label{w_1-2}
\ee
In this expansion, the first order term corresponds to the mean field theory, and higher order terms arise due to the correlation effects. $\delta \Upsilon(\rv,\rv)$ is defined as the correlation potential, and is related to the electrostatic Green's function $\mathcal{G}(\rv,\rv')$ via
\begin{subequations}
\label{SCFT-0}
\be
\delta \Upsilon(\rv,\rv) = \lim_{\rv' \rightarrow \rv}
 \left[
\mathcal{G} (\rv,\rv')
- \lim_{\rv'' \rightarrow \infty}
\mathcal{G} (\rv + \rv'',\rv'+ \rv'')
\right].
\ee
The Green's function is defined as incremental potential at $\rv$ due to a mono valence test ion $q$ inserted at $\rv'$, in the presence of the background potential $\Phi(\rv)$.
Substituting the preceding two equations back into (\ref{eq:Poisson}) and neglecting terms of higher order in $k$, we arrive at a {\em modified Poisson-Boltzmann equation}:
\ba
\label{eq:mpbe}
- \epsilon \nabla^2 \Phi(\rv) &=&
m q \rho^0_+ \,  e^{- \beta m q \Phi(\rv)
 - \frac{1}{2} m^2 \beta q \, \delta \Upsilon(\rv, \rv)}
\nonumber\\
&-& n q \rho^0_- \, e^{\beta n  q \Phi(\rv)
 - \frac{1}{2} n^2 \beta q \, \delta \Upsilon(\rv, \rv)}.
\ea
For details, see the Sec.~I of \worktwo.

To obtain a close system of equations, we need another equation for the Green's function $\mathcal{G} (\rv,\rv')$.  A self-consistent treatment is to consider $\mathcal{G} (\rv,\rv')$ as a linear perturbation to the background $\Phi(\rv)$ in  Eq.~(\ref{eq:mpbe}) and linearize in terms of $\mathcal{G} (\rv,\rv')$.  This leads to
\vspace{3mm}
 \ba
\label{eq:Greens-0}
- \epsilon \nabla^2 \mathcal{G}(\rv,\rv') &=&
- \beta q^2
\left[
m^2 \rho^0_+ \,  e^{- \beta m q \Phi(\rv)
 - \frac{1}{2} m^2 \beta q \, \delta \Upsilon(\rv, \rv)}
\right.
\nonumber\\
&& \left. n^2  \rho^0_- \, e^{\beta n  q \Phi(\rv)
 - \frac{1}{2} n^2 \beta q \, \delta \Upsilon(\rv, \rv)}
 \right]
\mathcal{G}(\rv,\rv')
\nonumber\\
&& + q \, \delta(\rv - \rv').
\ea
\end{subequations}
Three equations Eqs.~(\ref{SCFT-0}) form the {\em self consistent field theory} (SCFT) approximation. It has been studied by various authors for symmetric case $m = n$. \cite{wang,netz-variational,buyukdagli}  It is a more refined approximation than the classical Poisson-Boltzmann theory.

Let us define the (bare) Debye length $\debye$ and the Bjerrum length $b$ via:
\begin{subequations}
\ba
\debye = 1/\kappa &\equiv& \sqrt{\epsilon/  \beta q^2
( m^2 \rho^0_+ + n^2 \rho^0_- )},
\label{Debye_def} \\
b &\equiv& {\beta q^2}/{4 \pi \epsilon}.
\label{Bjerrum_def}
\ea
\end{subequations}
Same as in \worktwo, we shall measure all lengths in units of $\debye$, and define the dimensionless versions of mean potential $\Psi$ and Green's function $G(\rv,\rv')$ via
\begin{subequations}
\ba
&& \rv \rightarrow \rv \, \ell_{\rm DB},
\\
&&\Psi \equiv q\beta\Phi,
\label{eq:Psi}
\\
&&G \equiv q\beta\mathcal{G}.
\ea
\label{dimensionless_def}
\end{subequations}
For details, see Eqs.~(2.3) of \worktwo.  Eq.~(\ref{eq:mpbe}) and (\ref{eq:Greens-0}) then reduce to the following dimensionless form:
\begin{subequations}
\begin{widetext}
\vspace{-3mm}
\label{SCFT-eqn}
\ba
&&-\frac{d^2\Psi(z)}{dz^2}
+\frac{1}{m \! + \! n}
\! \left[
e^{n\Psi(z)-n^2 \Delta\varepsilon(z)}
-e^{-m\Psi(z)-m^2 \Delta\varepsilon(z)}
\right]
=  0,
\label{mPBE2}
\\
&& \left[ -\nabla^2 + \frac{1}{ m \! +\! n} \!
 \left[
 m  e^{ -m \Psi(z) - m^2 \Delta \varepsilon(z) }
+n e^{n \phi(z) - n^2 \Delta \varepsilon(z) }
\right]
\right]
\times G(\rv,\rv')
= g \, \delta(\rv-\rv'),
\label{ce mgf}
\ea
\end{widetext}
where $z$ is the distance to the charged plate, and $g = 4 \pi b/\ell_{\rm DB}$ a smaller parameter for a dilute electrolyte, and $\Delta \varepsilon(z)$ the correlation energy and is related to the electrostatic Green's function $G(\rv,\rv')$ via:
\be
\Delta \varepsilon(z)  =
\frac{1}{2} \lim_{\rv' \rightarrow \rv} \!
\Big[
G(\rv,\rv') - \lim_{\rv'' \rightarrow \infty}
{G} (\rv \! + \! \rv'',\rv' \! + \! \rv'')
\Big].
\label{mce}
\ee
\end{subequations}
We shall solve Eqs.~(\ref{SCFT-eqn}) perturbatively to the first order in $g$ in this work.  The leading order far field asymptotics of the mean potential has the following simple form:
\be
\Psi(z) = \eta_R \, e^{-\alpha z},
\ee
where $\eta_R$ is the renormalized surface charge density, given by Eq.~(\ref{eta_R_mn}), whilst $\alpha = \kappa_R/\kappa$ is given by Eq.~(\ref{m:n alpha-2}), with $\kappa_R$ the renormalized inverse Debye length.


\vspace{-4mm}
\subsection{ Perturbative Expansion in $g$}

We shall solve Eqs.~(\ref{SCFT-eqn}) using perturbation method, treating $g = 4 \pi b/\ell_{\rm DB}$  as a control parameter.  That means we expand $\Psi(z)$ and $G(\rv,\rv')$ into asymptotic series of $g$, and solve the coefficients order by order.  Since the source in the RHS of Eq.~(\ref{ce mgf}) is linear in $g$, whereas Eq.~(\ref{mPBE2}) is formally independent of $g $, we expect that $\Psi(z)$ starts with zero-th order, whilst $G(\rv,\rv'), \Delta \varepsilon(z)$ start with first order:
\begin{subequations}
\label{PT-series}
\ba
\Psi(z) &=& \Psi_0(z) + g \, \Psi_1(z)
\quad \,\, + \cdots,
\label{Psi-expansion}
\\
G(\rv, \rv') &=&  \quad 0 \quad +  g\, G_1(\rv, \rv')
+ \cdots, \\
\Delta \varepsilon(z) &= &  \quad  0 \quad
+ g \, \Delta \varepsilon_1 (z)
\quad \! + \cdots.
\ea
\end{subequations}
Substituting these back into Eqs.~(\ref{SCFT-eqn}), we find that, to the zero-th order,
$\Psi_0(z)$ satisfies the nonlinear Poisson-Boltzmann equation (PBE):
\be
-  \Psi_0'' (z)
+\frac{1}{m+n}
\left(e^{n\Psi_0(z)}
-e^{-m\Psi_0(z)}\right)=0,
\label{original PBE}
\ee
which, for the one plate geometry, was solved for arbitrary integers $m,n$ using the method of asymptotic matching discussed in \workone.  To the order in $g$, the Green's function $G_1(\rv,\rv')$ can be found in terms of $\Psi_0(z)$ by solving the following linear PDE:
\ba
&& \left[  -\nabla^2  +  \frac{1}{m+n}
\left(
n \, e^{-n \Psi_0(z)} + m \, e^{m \Psi_0(z)}
\right)
\right]
G_1(\rv,\rv')
\nonumber\\
&=& \delta(\rv -  \rv').
\label{ce gf}
\ea
From $G_1(\rv,\rv')$ we can obtain the first order correlation energy $\Delta \varepsilon_1(z)$ using Eq.~(\ref{mce}).  This problem has been solved in \worktwo, again for arbitrary $m,n$~\footnote{Only near field and far field asymptotics have been found for cases other than $1:-1$, $2:-1$, and $1:-2$. }.  The first order correction to potential, $\Psi_1(z)$, satisfies the following inhomogeneous linear ODE:
\ba
-  \Psi_1'' (z)
 +\frac{1}{m \! + \! n} \!
\left[ n \, e^{n\Psi_0(z)}
+m \, e^{-m\Psi_0(z)}\right] \!
\Psi_1(z)
=  S(z),
\nonumber\\
\label{phi_1-1:1}
\ea
where the source $S(z)$ is defined as
\be
S(z) \equiv \frac{1}{m+n}
\left(n^2 e^{n\Psi_0(z)}
-m^2 e^{-m\Psi_0(z)}\right)
\Delta \varepsilon_1(z).
\label{S_z-def}
\ee
and can be obtained in terms of $\Psi_0(z)$ and $\Delta \varepsilon_1(z)$.
Here we shall try to find $\Psi_1(z)$, for arbitrary valences $m,n$.

As we have shown in \workone, the solution to Eq.~(\ref{original PBE}) can be expressed in terms of a function $\Upsilon_{m,n}$ that depends on two integers $m,n$:
\be
\Psi(z) = \Upsilon_{m,n}(z + z_0).
\ee
The parameter $z_0$ shall be determined by enforcing the boundary condition Eq.~(\ref{original bc-0}).The function $\Upsilon(z)$ diverges logarithmically at $z = 0$.  As a consequence, the parameter $z_0$ goes to zero in the limit of infinite surface charge density.  It therefore can be treated as a small parameter for a strongly charged surface.

\subsection{The Boundary Conditions}
Same as in \workone \, and \worktwo, we shall take the convention that the plate is {\em positively charged}, so that the negative ions (with charge $-n e$) are the counter-ions and the positive ions (with charge $me$) are the co-ions.  In \workone \, and \worktwo, the coordinate system was chosen such that a plate with dimensionless surface charge density $\eta$ is located at $- z_0$, with $z_0$ chosen as a function of $\eta$, such that the potential $\Psi_0(z)$ is independent of the (dimensionless) surface charge density $\eta$, and diverges at $z = 0$.  This choice substantially simplifies the analyses in papers  I and I\!I.  In the present work, we shall choose a different coordinate system.  Namely we shall fix the plate at the origin $z = 0$.  It is then understood that all results in I and I\!I need to be transformed via $z \rightarrow z + z_0$ before they can be used in here.

The boundary conditions satisfied by the mean potential $\Psi(z)$ are given by (also in their dimensionless forms):
\begin{subequations}
\label{original bc}
\ba
\Psi'(0)
&=& -\eta, \quad \mbox{in the bulk},
\label{original bc-0}
\\
\Psi(\infty) &=& 0, \quad \mbox{on the plate},
\label{original bc-infty}
\ea
\end{subequations}
where
\be
\eta =  \frac{q\beta\sigma \ell_{\rm DB} }{\epsilon}
 = \frac{2 \ell_{\rm DB}} {\mu}
 \label{eta-def}
 \ee
is the dimensionless surface charge density. In writing Eqs.~(\ref{original bc-0}), we have assumed that the potential is constant to the left of the interface.

What we need are however the boundary conditions for $\Psi_0(z)$ and $\Psi_1(z)$ respectively.  It seems completely natural to require that both $\Psi_0(z)$ and $\Psi_1(z)$ vanishes as $z = \infty$.  But their boundary conditions at $z = 0$ are more subtle.  Eq.~(\ref{original bc-0}) only fix the boundary condition for the whole series.  It is conventional (and indeed seems very appealing) to require that $\Psi_0, \Psi_1,\ldots$ are all {\em independent of} $g$, and to expand both sides of Eq.~(\ref{original bc-0}):
\be
\Psi'_0(0) + g \, \Psi'_1(0) + \cdots
=  \eta + g \times 0 + \cdots.
\label{Psi'-decomp}
\ee
We can then enforce equality to hold order by order, and obtain a inhomogeneous boundary conditions for $\Psi_0(z)$ and a homogeneous one for $\Psi_1(z)$:
\begin{subequations}
\ba
\Psi_0'(0) &=& - \eta,
\quad
\Psi_0(\infty)=0;
\label{BC-phi0}
\\
\Psi_1'(0) &=&  0 ,
\quad\quad
\Psi_1(\infty)=0.
\label{BC-phi1}
\ea
 \label{BC-phi_0-phi_1}
\end{subequations}
We must remember, however, that Eqs.~(\ref{BC-phi0}) is only one of infinite number of possible choices.  In particular, the functions $\Psi_0(z), \Psi_1(z),\ldots$ need not to be independent of $g$.  In fact we can freely add a part to $g\, \Psi_1(z)$ and subtract it off from  $\Psi_0(z)$, such that Eq.~(\ref{Psi'-decomp}) is unaltered.  This subtle point provides the key to understand our renormalization group analysis below.



\subsection{Formal Solution to $\Psi_1(z)$}
\label{sec:H}
In order to solve Eq.~(\ref{phi_1-1:1}), we only need to find the corresponding Green's function $H(z,z')$, defined as:
\ba
&-& \!\! \frac{d^2 } {dz^2} H(z,z')
 +\frac{1}{m \! + \! n} \!
\left[ n \, e^{n\Psi_0(z)}
+ m \, e^{-m\Psi_0(z)}\right]  \!\!
H(z,z')
\nonumber\\
&=&  \delta(z-z'),
\label{Green's function}
\ea
together with homogeneous boundary conditions at $z = 0$ and at $z = \infty$.  Note that $H(z,z')$ is a one dimensional Green's function, whilst $G_1(\rv,\rv')$ in Eq.~(\ref{ce gf}) is a three dimensional Green's function.  As is well known, $H(z,z')$ can be constructed using the standard Liouville method \cite{stone-goldbart}.  For this purpose, we need two independent homogeneous solutions $\phi_L(z)$ and $\phi_R(z)$ to  Eq.~(\ref{Green's function})~\footnote{The the subscripts ``L'' and ``R'' refer to left and right respectively. }:
\be
- \phi''_{L,R}(z)
 +\frac{1}{m \! + \!n} \!
\left(ne^{n\Psi_0(z)}
+me^{-m\Psi_0(z)}\right) \phi_{L,R}(z)
=  0,  \label{phih-eqn}
\ee
subjected to the homogeneous boundary conditions
\begin{subequations}
\ba
\phi_L'(0) &=& 0,
\label{phi_L-BC}
\\
\phi_R(\infty) &=& 0.
\label{phi_R-BC}
\ea
\label{phi_LR-BC}
\end{subequations}
Taking the derivative of the original PB equation (\ref{original PBE}) with respect to $z$, we find that $\Psi'_0(z)$ satisfies Eq.~(\ref{phih-eqn}).  Furthermore, since $\Psi_0(z)$ decays as $e^{-z}$ for $z \gg 1$, so does its derivative.   Therefore $- \Psi'_0(z)$ is precisely the homogeneous solution $\phi_R(z)$ that we are looking for:
\be
\phi_R(z) \equiv - \Psi'_0(z).
\label{phi_R-def}
\ee
The other solution $\phi_L(z)$ can be obtained by the method of variation of parameters.  Let
\be
\phi_L(z) = f(z)\phi_R(z),
\label{phi_L-f}
\ee
and substituting it back into Eq.~(\ref{phih-eqn}), we find $f(z)$ satisfies the following equation:
\be
\frac{d^2f(z)}{dz^2}+2\frac{df(z)}{dz}\frac{d\phi_R(z)}{dz}=0.
\ee
This equation can be readily solved:
\be
f(z) = \int{\frac{dz}{\phi_R(z)^2}} + f_1.
\label{vop}
\ee
Hence
\be
\phi_L(z) = \phi_R(z) \left( \int{\frac{dz}{\phi_R(z)^2}} + f_1 \right).
\label{phi_L-phi_R}
\ee
The constant $f_1$ shall be determined by the boundary condition satisfied by $\phi_L$ at $z = 0$, Eq.~(\ref{phi_L-BC}).

The Wronskian formed by two functions $g(z)$ and $h(z)$ is defined as
\be
W(g, h; z) = g(z)h'(z) - h(z) g'(z).
\label{wronskian}
\ee
Using Eq.~(\ref{phi_L-phi_R}), it can be easily shown that
\be
W(\phi_L, \phi_R; z) = -1.
\label{Wronskian-LR}
\ee

The Green's function $H(z,z')$ can now be obtained:
\ba
 H(z, z^\prime) &=&
\left\{ \begin{array}{ll}
{\displaystyle
 \phi_L(z)\phi_R(z'),
 }
  &
   \quad \mbox{($z < z^\prime$)},
   \vspace{3mm}\\
{\displaystyle
\phi_L(z') \phi_R(z),
 }
 &
   \quad \mbox{($ z > z^\prime$)}.
   \end{array}  \right.
\label{GF1:1}
\ea

The first order correction $\Psi_1(z)$ can be now expressed in terms of the Green's function as
\begin{subequations}
\ba
\Psi_1(z)
&=&
\int_0^{\infty} H(z,z') S(z') dz.
\nonumber\\
&=& \phi_R(z)\int_{0}^{z}\phi_L(z')S(z')dz'
\nonumber\\
&+& \phi_L(z)\int_{z}^{\infty}\phi_R(z')S(z')dz'.
\label{phi-gr-general}
\ea
\end{subequations}
Constructed as such, $\Psi_1(z)$ naturally satisfies the homogeneous boundary conditions Eqs.~(\ref{BC-phi1}) at both ends.  In later sections, we shall use this general expression and previous results of $\Psi_0(z)$ and $\Delta \varepsilon_1(z)$ to calculate $\Psi_1(z)$ for the generic values of $m, n$.


\subsection{Subtleties of the Correlation Energy}
In \worktwo, it was shown that the first order correlation energy $\Delta \varepsilon(z)$ can be decomposed into two parts:
\be
\Delta \varepsilon_1(z) = \Delta \varepsilon^{\infty}(z)
+ \delta \varepsilon(z).
\ee
Here $\Delta \varepsilon^{\infty}(z)$ scales as $- 3 g / 16 \pi (z + z_0)$ in the near field, and decays exponentially in the far field.  Furthermore, it is manifestly independent of the dielectric constant of the plate $\epsilon_1$.  By contrast, the second part $\delta \varepsilon(z)$ depends on the dielectric constant of the plate $\epsilon_1$, but is subdominant to $\Delta \varepsilon^{\infty}(z)$ except in a very thin region close to the plate $z \ll z_0 \sim \mu$, where $\mu$ is the Gouy-Chapman length.  This regime is called the {\em extremely near field} in \worktwo.  Evidently, $\delta \varepsilon(z)$ becomes important in this regime because of the image charge effects due to the discontinuity of dielectric constant on the interface.  This effects is screened by the counter-ions once the test-ion is couple of $\mu$ away from the plate.

If the dielectric constant $\epsilon_1$ of the plate is smaller than $\epsilon$ that of the solvent (as is the usual case of insulator plate inside aqueous solvent), $\delta \varepsilon(z)$ diverges to $+ \infty$ as $z \rightarrow 0^+$, that is, as the test-ion approaches the plate.  The effect of this repulsive image charge is to push the test ion a few $\mu$ away from the plate.  This effects can be largely ignored in our calculation of $\Psi_1(z)$, as long as $\mu \ll \ell_{\rm DB}$.  By contrast, if $\epsilon_1 > \epsilon$ (as in the case of conductor plate), $\delta \varepsilon(z)$ diverges to $- \infty$ as $z \rightarrow 0^+$, and the image charge strongly attract the test ion in the extremely near field.  This attraction will have major influence on the statistical distribution of counter-ions and therefore can not be neglected in our calculation of $\Psi_1(z)$.

In this work, we shall always assume $\epsilon_1 < \epsilon$, and hence the correction $\delta \varepsilon(z)$ can be safely ignored.  This substantially simplify our analyses below.


\section{Symmetric Electrolyte}
\label{sec:symmetric}
Let us apply the general formalism developed above to the simplest case of $1:-1$ symmetric electrolyte.  The solution to the PBE Eq.~(\ref{original PBE}) for the one-plate geometry is well known:
\ba
\Psi_0(z)=
2\log \coth \left( \frac{z+ z_0}{2}\right).
\label{phi0 1:1}
\ea
Note that $\Psi_0(z)$ diverges logarithmically at $z = - z_0$, where the parameter $z_0$ is small for a strongly charged plate, and remains undetermined at this stage, In the far field, $\Psi_0(z)$ scales as
\be
\Psi_0(z) =  4\, e^{- (z + z_0)} + O(e^{- 2 z}),
\quad z\rightarrow \infty.
\label{1:1 phi0 far}
\ee

Recall that we neglect the part of the correlation energy that explicitly depends on the dielectric constant of the plate $\epsilon_1$.  The remaining part $\Delta \varepsilon^{\infty}(z)$ is independent of $\epsilon_1$, and is given by \cite{xing-bing}:
\begin{widetext}
\vspace{-4mm}
\ba
\Delta \varepsilon^{\infty}(z)
&=&
\frac{ e^{-2(z+ z_0) }}{16\pi ( z+ z_0 )}
-\frac{1}{16\pi}{\rm csch}^2(z+ z_0)
\big[
\log{4(z+ z_0)}+E_1(4 (z+ z_0))
+\gamma
\big], \nonumber
\ea
where
\be
E_1(z)=\int^{\infty}_{1}{t^{-1}e^{-tz}}dt
\ee
is one of the generalized exponential integral functions and $\gamma$ is the Euler constant. The near field and far field asymptotic behaviors of the correlation energy are
\ba
\Delta \varepsilon^{\infty}(z)
\sim \left\{
\begin{array}{ll}
\!\!{\displaystyle
-\frac{3}{16\pi (z+z_0)}, \quad
}
 &z \rightarrow 0;
\vspace{2mm}\\
\!\!{\displaystyle
-\frac{1}{4\pi}\!\left[ \gamma
+ \log{4(z \! +\!z_0)}
- \frac{1}{4(z\!+ \!z_0)}\right]
\!e^{-2(z+z_0)},
}
\quad &z \rightarrow \infty.
\end{array} \right.
\ea
\vspace{3mm}
\end{widetext}
All near field asymptotics are valid only in the strongly charged regime where $z_0 \ll 1$.  The asymptotic behaviors of the source term $S(z)$ (c.f. Eq.~(\ref{S_z-def}), and with $\Delta \varepsilon(z)$ approximated by $\Delta \varepsilon^{\infty}(z)$) are given by:
\ba
S(z)\sim\left\{ \begin{array}{ll}
{\displaystyle
 -\frac{3}{8\pi(z+z_0)^3},
}
 \quad &z \rightarrow 0;
 \vspace{2mm}\\
{\displaystyle
-  \frac{1}{4\pi}
 \Big(\gamma
 + \log{4(z+z_0)}
  \Big)
  e^{-2(z+z_0)} ,
  }
 \quad
 &z \rightarrow \infty.
\end{array} \right.
\label{1:1 S(z)}
\ea
Two homogeneous solutions $\phi_{L}(z), \phi_{R}(z)$ to Eq.~(\ref{phih-eqn}) can also be easily found using Eqs.~(\ref{phi_R-def}) and (\ref{phi_L-phi_R}):
\ba
\phi_R(z) &=& -  \frac{d\Psi_0(z)}{dz}
= 2 \, {\rm csch}(z+z_0),
\\
\phi_L (z) &=& 2\,
{\rm csch}(z+z_0)
\Big[
- \frac{1}{8} \, ( z + z_0 )
\nonumber\\
&+& \frac{1}{16}\, \sinh{ 2 ( z + z_0 ) }
 + f_1(z_0)
\Big],
\ea
where the function $f_1(z_0)$ is fixed by the boundary condition Eq.~(\ref{phi_L-BC}):
\ba
f_1(z_0) &=& \frac{z_0}{8}
+ \frac{1}{16}\, \sinh(2z_0)
- \frac{1}{4} \,\tanh(z_0)
\nonumber\\
&=& \frac{z_0^3}{6} + O(z_0^5).
\label{C_L 1:1}
\ea
The leading order near field asymptotics of $\phi_L, \phi_R$ are
\begin{subequations}
\label{near-asymp-phi_LR}
\ba
\phi_R(z) &\sim& \frac{2}{(z+z_0)} ,
\\
\phi_L(z) &\sim&  \frac{z_0^3}{3(z+z_0)}
+  \frac{(z+z_0)^2}{6} ,
\ea
\end{subequations}
whilst their far field asymptotics are
\begin{subequations}
\label{far-asymp-phi_LR}
\ba
\phi_R(z) &\sim& 4\, e^{- (z  + z_0)}, \\
\phi_L(z) &\sim&  \frac{1}{8} \, e^{z+z_0}.
\ea
\end{subequations}
The first order correction to the mean potential $\Psi_1(z)$ is then given by Eq.~(\ref{phi-gr-general}) with $\phi_{L,R}(z)$ and $S(z)$ given by the above results.


\subsection{Near-field and Far-field Behaviors}
To determine the parameter $z_0$, we impose the boundary conditions Eqs.~(\ref{BC-phi_0-phi_1}).    Using of Eq.~(\ref{phi0 1:1}) in Eq.~(\ref{BC-phi0}) leads to:
\ba
 z_0 &=& \rm{ArcSinh} (2 \, \eta^{-1})
  \label{z_0:eta-sym} \\
&=& {2}\, {\eta}^{-1} + O(\eta ^ {-2}). \nonumber
\ea
For a strongly charged plate, $z_0$ is a small number.

Using the following identity:
\be
\int_0^z dz' = \int_0^{\infty} dz' - \int_z^{\infty} dz' ,
\ee
in Eq.~(\ref{phi-gr-general}), we can rewrite the first order perturbation solution to $\Psi(z)$ in the following form:
\ba
\Psi(z)
&=& \Psi_0(z) + g \, \Psi_1(z)
\nonumber\\
&=&
 \Psi_0(z) + g\, \phi_R(z)
 \int_{0}^{\infty}\phi_L(z')S(z')dz'
  \nonumber \\
&+& g\, \bigg(\phi_L(z)\int_{z}^{\infty}
	\phi_R(z')S(z')dz'
\nonumber\\
&-& \phi_R(z)\int_{z}^{\infty}
	\phi_L(z')S(z')dz' \bigg).
\label{phi_1_sym}
\ea
Using the far field asymptotics Eqs.~(\ref{1:1 S(z)}) and Eqs.~(\ref{far-asymp-phi_LR}), we easy see that
each of three integrals in Eq.~(\ref{phi_1_sym}) converges separately.  Furthermore, the first two terms scale as $e^{-z}$, whereas the last two terms (inside the bracket) scale as $\log(z) \, e^{-2 z}$ for large $z$. Therefore the latter does not contribute to the leading order far field asymptotics of $\Psi(z)$.  Using Eq.~(\ref{1:1 phi0 far}) and (\ref{far-asymp-phi_LR}), we obtain the following leading order far field asymptotics of $\Psi(z)$:
\ba
\Psi(z) &=& 4\,  \left[
1 + g\, \int_{0}^{\infty}\phi_L(z')S(z')dz'
\right]  \,  e^{- z- z_0}
\nonumber\\
&+&  O\Big( g\, \log(z) \,e^{-2z}\Big).
\label{phi1 1:1 expression}
\ea

We still need to calculate the integral in Eq.~(\ref{phi1 1:1 expression}), which, even though remains finite for arbitrary finite $z_0$, nevertheless becomes singular in the strongly charged limit, i.e. $z_0 \sim 2/\eta \rightarrow 0$.  To see this, let us analyze the near field asymptotics of the integrand.  Using Eq.~(\ref{1:1 S(z)}) and Eq.~(\ref{near-asymp-phi_LR}), we find that for $z, z_0 \ll 1$:
\be
\phi_L(z) S(z) = - \frac{1}{16 \pi (z+z_0)}
- \frac{z_0^3}{8 \pi(z+z_0)^4} + O(1).
\ee
As $z_0 \rightarrow 0$, integration of the first term gives $(16 \pi)^{-1} \log z_0$, whereas that of the second term gives a finite number.  Therefore the following limit exists:
\ba
C_s \equiv \lim_{z_0\rightarrow 0}\left(\int_{0}^{\infty}\phi_L(z')S(z')dz'
- \frac{1}{16 \pi}\log{z_0} \right).
\label{1_1 Cs}
\ea
Numerical integration using Wolfram Mathematica gives
\be
C_s \approx 0.005673.
\label{C_s-11}
\ee
Using this result in Eq.~(\ref{phi1 1:1 expression}), and using Eq.~(\ref{z_0:eta-sym}) to trade $z_0$ in for $\eta$, we finally obtain the leading order far field asymptotics of $\Psi(z)$ (in the strongly charged limit):
\ba
\Psi(z) &=&
 {4\left(  1 - \frac{g}{16 \pi}
 \log \left( \frac {\eta} {2} \right) + g \,C_s
 + O(\eta^{-1} )
 \right) }
 \, e^{-z}
\nonumber\\
 &+& O\left( g\,\log (z) \, e^{-2 z} \right),
\label{1:1 final result}
\ea
The coefficient of $e^{- z}$ defines the {\em renormalized surface charge density} $\eta_R(\eta, g)$ of a highly charged surface, calculated to the leading orders in $g$ and in $\eta$:
\be
\eta_R(\eta, g) = 4 - 4\,g \Big[ \frac{1}{16\pi}
 \log \left( \frac{\eta} {2} \right)  -  C_s
 \Big]
  + O(\eta^{-1})  + O(g^2).
  \label{eta_R_11}
\ee
In the dilute and strongly charged limit, $g \rightarrow 0, \eta \rightarrow \infty$, and $\eta_R \rightarrow 4$, which is what we obtained in \workone.

Lau \cite{lau} studied the one-loop correction to surface charge density of an infinitely thin charged plate inside $1:-1$ electrolyte.  We note that boundary conditions used by Lau are different from ours.



\section{m:-n Asymmetric Electrolyte}
\label{sec:asymmetric}

For the generic case of $m:-n$ electrolyte, there is no closed form for the zero-th order solution $\Psi_0(z)$ (except for the cases of $2:-1$ and $1:-2$).  Nevertheless, we can find both near field and far field expansions up to arbitrary orders.  As is shown in Eqs.~(22), (30) of \workone \,  \footnote{Recall that the $z+z_0$ in this work corresponds to $z$ in \workone.}, the leading order near field and far field asymptotics of $\Psi_0(z)$ are given by
\ba
\Psi_0(z) &=& \Upsilon_{m,n}(z + z_0)
\nonumber\\
&\sim&
\left\{
\begin{array}{ll}
{\displaystyle
\frac{1}{n}
\log{\frac{2(m+n)}{n(z+z_0)^2}},}
\quad &
z \rightarrow 0;
\vspace{3mm}
\\
{\displaystyle
c_1^{m,n} \, e^{-(z+z_0)} ,
}
&
z \rightarrow \infty,
\end{array}
\right.
\label{Psi_0-mn}
\ea
wherex $\Upsilon_{m,n}(w)$ is a universal function that only depends on two integers $m, n$.
Note that the near field asymptotics is valid only for strongly charged plates, for which $z_0 \ll 1$.  The numerical values of $c_1^{m,n}$ was calculated and tabulated for many cases in \workone.   Unlike the case of symmetric electrolytes, however, here we shall {\em not} impose the boundary conditions Eqs.~(\ref{BC-phi_0-phi_1}). Instead, we shall first obtain a globally well-behaved approximation for the mean potential $\Psi(z)$, and then determine the value of $z_0$ by imposing Eq.~(\ref{original bc}).

The near/far field asymptotics of the correlation energy are given by Eqs.~(5.14a) and (5.26) in \workone:
\ba
\Delta \varepsilon_1(z)
\sim
\left\{
\begin{array}{ll}
{\displaystyle
-\frac{3}{16\pi (z+z_0)},
}
&
z \rightarrow 0;
\vspace{3mm}
\\
{\displaystyle
\frac{(\log3)(m-n) c_1^{m,n}}{16\pi}
\, e^{-(z+z_0)} ,
}
\quad &
z \rightarrow \infty.
\end{array}
\right.
\ea
The function $S(z)$ is related to $\Psi_0(z)$ and $\Delta \varepsilon_1(z)$ via Eq.~(\ref{S_z-def}).
Its far field and near field asymptotics are:
\be
S(z) \equiv
\mathcal{S}(z + z_0)
\sim \left\{
\begin{array}{ll}
{\displaystyle
-\frac{3n}{8\pi(z+z_0)^3},
}
\quad
&
z \rightarrow 0;
\vspace{3mm}
\\
{\displaystyle
-  s_{m,n}\,
c_1^{m,n} \,e^{-(z+z_0)} ,
}
&
z \rightarrow \infty.
\end{array}
\right.
\label{S-mn-asympt}
\ee
where
\be
s_{m,n}\equiv\frac{\log{3}}{16\, \pi} \,(m-n)^2.
\label{s_mn-def}
\ee
For symmetric electrolyte, $s_{m,n} = 0$, and $S(z)$ scales as $e^{-2 z}$ in the far field.

Two homogeneous solutions $\phi_R,\phi_L$  to Eq.~(\ref{phih-eqn}) were already formally constructed in Eqs.~(\ref{phi_R-def}), (\ref{phi_L-phi_R}).  The factor $f_1$ in Eq.~(\ref{phi_L-phi_R}) depends on the parameter $z_0$, and can be found by imposing the boundary condition Eq.~(\ref{BC-phi1}).  Using Eq.~(\ref{Psi_0-mn}), (\ref{phi_R-def}), (\ref{phi_L-phi_R}), we determine the leading order near field asymptotics of $\phi_{L,R}(z)$,
\begin{subequations}
\ba
\phi_R(z) &\sim& \frac{2}{n(z+z_0)} + O(z+z_0),
\label{phi_R-mn-near}
\\
\phi_L(z) &\sim&  \frac{2 f_1(z_0)}{n(z+z_0)}
+  \frac{n}{6}\, (z+z_0)^2.
\label{phi_L-mn-near}
\ea
\label{phi-mn-near}
\end{subequations}
Now imposing the boundary condition Eq.~(\ref{phi_L-BC}) on Eq.~(\ref{phi_L-mn-near}), we find (c.f. Eq.~(\ref{C_L 1:1}) for the $1:-1$ case):
\be
f_1(z_0) \sim \frac{n^2}{6} z_0^3 + O(z_0^5).
\label{f_1-mn}
\ee

We can also obtain the far-field asymptotics of $\phi_R(z)$:
\begin{subequations}
\label{phi-mn-far}
\ba
\phi_R(z) = - \Psi'_0(z) \sim c_1^{m,n} \, e^{-(z+z_0)},
\label{phi_R-mn-far}
\ea
Combining this with Eq.~(\ref{Wronskian-LR}), we obtain the leading order far field asymptotics of $\phi_L(z)$:
\ba
\phi_L(z) \sim  \frac{1}{2c_1^{m,n}} \, e^{(z+z_0)}.
\label{phi_L-mn-far}
\ea
\end{subequations}
Note that the part $f_1(z_0)$ in Eq.~(\ref{phi_L-phi_R}) does not contribute to the leading order far field asymptotics of $\phi_L(z)$.

\subsection{First Order Correction and Secular Term}

The first order correction Eq.~(\ref{phi-gr-general}) is repeated here:
\ba
\Psi_1(z) &=&
\phi_R(z)\int_{0}^{z}\phi_L(z')S(z')dz'
\nonumber\\
&+& \phi_L(z)\int_{z}^{\infty}\phi_R(z')S(z')dz'.
\label{phi-gr-m:n}
\ea
In the far field, $z \gg 1$, all functions in the second term can be replaced by their leading order far field asymptotics,  i.e., Eqs.~(\ref{phi-mn-far}) and Eq.~(\ref{S-mn-asympt}).  The integral then becomes trivial:
\be
\phi_L(z)\int_{z}^{\infty}\phi_R(z')S(z')dz'
\sim -\frac{1}{4}c_1^{m,n} \,  s_{m,n}\, e^{-(z+z_0)}.
\ee
By the same token, we can also replace $\phi_R(z)$ in front of the first integral in Eq.~(\ref{phi-gr-m:n}) by its far field asymptotics.  This leads to the following asymptotics for $\Psi(z)$ (up to the order of $g$) in the far field regime:
\ba
\Psi(z) &=& \Psi_0(z) + g \, \Psi_1(z)
\label{phi-gr-m:n-2}\\
&\sim&
\left[  1 \! -\! \frac{1}{4} g\,  s_{m,n}
 \!+\! g \!\! \int_{0}^{z} \!\! \phi_L(z')S(z')dz'
\right]\! c_1^{m,n} \, e^{- (z+z_0)}.
\nonumber
\ea
Inside the bracket, the first term (independent of $g$) comes from the nonlinear PB theory, whereas the other two terms (both linear in $g$) come from the electrostatic correlations.

We still need to calculate the remaining integral in Eq.~(\ref{phi-gr-m:n-2}).  Let us first introduce a sufficiently large number $z^*$ so that for $z' > z^*$, we can use far-field asymptotics for $\phi_L(z')$ and $S(z')$, Eqs.~(\ref{phi-mn-far}), and (\ref{S-mn-asympt}).  The portion of integral from $z^*$ to $z$ can then be approximately calculated:
\ba
g\, \int_{z^*}^{z}\phi_L(z')S(z')dz'
\sim
-\frac{1}{2} g\, s_{m,n} \, (z-z^*).
\label{secular-term-mn}
\ea
The integral therefore {\em grows linearly with $z$ without bound} as $z \rightarrow \infty$.  Substituting this back into Eq.~(\ref{phi-gr-m:n-2}), we see that the correction due to electrostatic correlations becomes much larger than $\Psi_0(z)$, the mean field potential predicted by PB, for sufficiently large $z$.  Such a perturbative correction is usually called a {\it secular term} and indicates the breakdown of regular perturbation method, in the regime $g\,s_{m,n}\,z \geq 1$.  A perturbation problem with secular term is called a {\em singular perturbation problem}.

\subsection{Renormalization Group (RG) Method}
There are many kinds of singular perturbation problems, and there seems no existing universal method capable of dealing with all problems.  Heuristically speaking, the reason underlying this unsatisfactory {\em status quo} is that regular perturbation method may break down in many different ways, and discovery of the most relevant method is often led by an intuitive understanding of the particular problem.

Let us look at the ODE satisfied by $\Psi_1(z)$, Eq.~(\ref{phi_1-1:1}), in the far field regime, where $\Psi_0(z)$ can be set to zero:
\be
- \Psi''_1(z) + \Psi_1(z) = S(z).
\label{Psi_1-S-mn}
\ee
It has two homogeneous solutions $e^{\pm z}$.  Now for asymmetric electrolytes $m \neq n$, the source term, given by Eq.~(\ref{S-mn-asympt}), scales as $e^{-z}$, which is {\em proportional to one of the two homogeneous solutions in the far field}.  If Eq.~(\ref{Psi_1-S-mn}) is viewed as a linear system, then there is resonance between the {\em input} $S(z)$ and the {\em output} $\Psi_1(z)$, and the amplitude of the output is expected to grow linearly with $z$.  This is exactly what we see Eq.~(\ref{secular-term-mn})!  Such a resonance phenomenon is rather common in  many singular perturbation problems \cite{Bender-Orszag}, such as Duffing equation, Rayleigh equation etc..  As is well known, in these problems, the appearance of secular terms suggests the existence of slowing evolving variables that renormalize the characteristic time/length scales of the systems \cite{goldenfeld}.  In our case, we expect that the correlation energy renormalizes the Debye length so that it is no longer given by Eq.~(\ref{Debye_def}), as predicted by linearized PB.   As a consequence, in the dimensionless form, the average potential should decay as $e^{- \alpha z}$  where $\alpha = 1 + O(g)$.  Blind expansion of this function $e^{- \alpha z}$ in terms of $g$ would give us the secular term $- \alpha z$, as we have obtained via a mechanical perturbation analysis.  The method of renormalization group (RG) transformation is ideal for summing up all these secular terms and obtaining sensible results that are valid for all $z$.

Let us now come back to the issue of boundary conditions.  In Sec.~\ref{sec:H} we constructed the Green's function $H(z,z')$ and hence $\Psi_1(z)$ such that they satisfy the homogeneous boundary condition at $z = 0$, Eq.~(\ref{BC-phi1}).  We are, however, perfectly allowed to relax these boundary conditions, and to add to $\Psi_1(z)$ an arbitrary homogeneous solution $C\,\phi_R(z)$. \footnote{The other homogeneous solution $\phi_L(z)$ can not be added, because it will spoil the boundary condition at $z = \infty$. }  We can therefore rewrite $\Psi(z)$ in the following form:
\ba
&& \Psi(z, z_0,C) \equiv \Psi_0(z,z_0)
 + g \Big( \Psi_1(z, z_0) +  C\, \phi_R(z,z_0) \Big)
\nonumber\\
&=& \Psi_0(z, z_0)
+ g \, \phi_L(z, z_0)\int_{z}^{\infty}
\phi_R(z', z_0)S(z', z_0)dz'
\nonumber\\
&+& g \,  \phi_R(z, z_0) \left(
C +  \int_{0}^{z}\phi_L(z', z_0)S(z', z_0)dz'
\right).
 \label{phi-gr-2:1-rg}
\ea
Note that we have explicitly shown the dependence of various functions on the parameter $z_0$ as well.  Among these, $\phi_R$ and $S$ depends on $z$ and $z_0$ only through the sum $z + z_0$, whereas $\phi_L$ depends on two variables in a non-additive way, see Eqs.~(\ref{phi_L-phi_R}) and (\ref{f_1-mn}).

The perturbative solution Eq.~(\ref{phi-gr-2:1-rg}) automatically satisfies the boundary condition at $z = \infty$, Eq.~(\ref{original bc-infty}), and we still need to impose the other BC, Eq.~(\ref{original bc-0}) at $z = 0$.  On the other hand, Eq.~(\ref{phi-gr-2:1-rg}) contains two arbitrary parameters $z_0, C$. These two parameters can not be truly independent of each other.  In another word, if we tune $C$ slightly, there must be a way to tune $z_0$ appropriately, such that the solution Eq.~(\ref{phi-gr-2:1-rg}) remains invariant.  \footnote{If we carry out the perturbation series up to infinite order, this would be an exact invariance.  At this stage,  however, we have only worked out the perturbation series to the first order, hence the invariance is valid only up to the order of $g$.  }   This consideration suggests the existence of a one-parameter family of solutions to the original problem, defined by Eqs.~(\ref{SCFT-eqn}), that are equivalent to each other.  This allows us to carry out a renormalization group transformation.


Let's vary $C$ and $z_0$ simultaneously such that the mean potential Eq.~(\ref{phi-gr-2:1-rg}) is invariant up to the order of $g$:
\vspace{2mm}
\ba
O(g^2) &=&  d\Psi(z,z_0,C)
= \frac{\partial \Psi}{\partial z_0}  d z_0
+  \frac{\partial \Psi}{\partial C} dC
\nonumber\\
&=&  \phi_R( z, z_0 ) \left( - d z_0 + g\, dC \right)
\nonumber\\
&+&  g\, \left(
\frac{\partial \Psi_1 }{\partial z_0}
+ C\, \frac{\partial \phi_R }{\partial z_0}
\right) dz_0,
\label{rgeq}
\ea
where we have used the following identities:
\ba
\frac{\partial \Psi_0}{\partial z_0}
=\frac{\partial \Psi_0}{\partial z}
= \frac{\partial \Psi}{\partial C}
= - \phi_R(z, z_0).
\ea
Applying the argument of dominant balance to Eq.~(\ref{rgeq}), we easily see that $dz_0 \sim g \, dC$, and hence the bracket in Eq.~(\ref{rgeq}), being linear in $g \, dz_0$, is of higher order in $g$ and therefore can be neglected, since we only keep terms of order $g$.  Consequently we find the following first order {\em renormalization group equation}:
\be
d z_0 = g \, dC.
\ee
Integrating once, we find the relation between $z_0$ and $C$:
\be
z_0(C) = g \, C + \bar{z}_0,
\label{rz_0}
\ee
where $\bar{z}_0$ is a constant to be determined later by boundary condition.  Therefore, replacing $z_0$ by $z_0(C)$ in Eq.~(\ref{phi-gr-2:1-rg}), we are guaranteed to obtain a one-parameter family of solutions (parameterized by $C$) that are equivalent to each other up to the order of $g$:
\begin{widetext}
\vspace{-2mm}
\ba
\Psi \left( z, z_0(C), C \right)
&=& \Psi_0( z, z_0(C) )
+ g \, \Psi_1(z, z_0(C))
+  g\,C \phi_R(z,z_0(C))
\nonumber\\
&=&  \Upsilon(z + z_0(C) )
\nonumber\\
&+& g \, \phi_L(z,  z_0(C) )
\int_{z}^{\infty} \Upsilon'(z' + z_0(C) )
\mathcal{S}(z' + z_0(C) ) dz',
\nonumber\\
&-&  g \,  \Upsilon'(z +z_0(C) ) \left(
C +  \int_{0}^{z}\phi_L(z', z_0(C) )
\mathcal{S}(z' +z_0(C) ) dz'
\right).
 \label{phi-gr-mn-rg-2}
\ea
\end{widetext}
\vspace{-3mm}

Now comes the most crucial step of RG  transformation.  We shall bootstrap the parameter $C$ to be a {\it function} of $z$, $C(z)$, such that the approximate solution Eq.~(\ref{phi-gr-mn-rg-2}) is free of secular term.   Comparing with Eq.~(\ref{secular-term-mn}) we easily see that the choice
\be
C(z) = \frac{1}{2} s_{m,n}\, z
\label{rc-mn}
\ee
fulfills this purpose.  Let us check this explicitly.  Upon the afore-mentioned replacement, the integral inside the bracket in Eq.~(\ref{phi-gr-mn-rg-2}) becomes
\be
 \int_{0}^{z} \phi_L \left( z',  g\,s_{m,n}\, z' /2 +\bar{z}_0 \right)
 S \left( z', g\,s_{m,n}\, z'/2 +\bar{z}_0 \right)dz'.
 \label{integral-secular-mn}
\ee
We can use the far field asymptotics for two functions, Eq.~(\ref{phi_L-mn-far}) and (\ref{S-mn-asympt}), in the integrand:
\ba
 \hspace{-5mm}  \phi_L(z', g\,s_{m,n}\, z'/2 + \bar{z}_0 )
 &\sim& \frac{1}{2 \,c_1} e^{\alpha z' + \bar{z}_0},
 \\
S(z', g\,s_{m,n}\, z' /2 + \bar{z}_0 )
 &\sim&- s_{m,n} c_1^{m,n} \, e^{-( \alpha z' + \bar{z}_0) },
 \nonumber\\
\ea
where
\be
\alpha = 1 + g \,s_{m,n} /2.
\label{m:n alpha}
\ee
Using these in the integral Eq.~(\ref{integral-secular-mn}), we find that it contains the following secular term:
\be
{\rm Integral} = - \frac{1 }{2} \,s_{m,n} \,z + {\rm finite},
\ee
which is exactly canceled by our choice of $C(z)$, Eq.~(\ref{rc-mn}).  In another word, we have proved that the following limit exist:
\ba
&& h_{m,n}(\bar{z}_0 ) \equiv
\lim_{z \rightarrow \infty}
 \Bigg[
\frac{1}{2}\,  s_{m,n} \,z
\\
&+&  \int_{0}^{z}\phi_L(z', g \,  s_{m,n} z/2 +\bar{z}_0 )
S(\alpha z' +\bar{z}_0 )dz'
\Bigg].
\nonumber\\
\label{h_mn-def}
\ea

Note that Eq.~(\ref{rc-mn}) is not the only way to remove the secular term.  In fact, there are an infinite number of choices that are equally good, characterized by one arbitrary constant $C_0$: $C(z) = \frac{1}{2} s_{m,n}\, z + C_0.$ We shall see below why the particular choice $C_0 = 0$ is the most convenient one.

Substituting Eq.~(\ref{rc-mn}) back into Eq.~(\ref{phi-gr-mn-rg-2}) we find the {\em renormalized}
average potential:
\begin{widetext}
\ba
\Psi^R(z, \bar{z}_0)
&=& \Psi(z, z_0( C(z) ), C(z) )
\nonumber\\
&=& \Psi_0( z, z_0(C(z)) )
+ g \, \Psi_1(z, z_0(C(z)) )
+ g\, C(z) \phi_R(z,z_0(C(z)))
\nonumber\\
&=& \Upsilon( \alpha z +\bar{z}_0 )
+ g \, \phi_L(z,  (\alpha - 1) z +\bar{z}_0 )
\int_{z}^{\infty}\phi_R(\alpha z' +\bar{z}_0 )
S(\alpha z'  +\bar{z}_0 )dz',
\nonumber\\
&+&  g \,  \phi_R( \alpha z'  +\bar{z}_0  ) \left[
\frac{1}{2}\,  s_{m,n} \,z
+  \int_{0}^{z}\phi_L(z', g \,  s_{m,n} z/2 +\bar{z}_0 )
S(\alpha z' +\bar{z}_0 )dz'
\right],
 \label{phi-gr-mn-rg-3}
\ea
\end{widetext}


\subsection{Renormalized Potential Solves Modified PBE}

It remains to be shown that the renormalized potential Eq.~(\ref{phi-gr-mn-rg-3})
is still an approximate solution to Eq.~(\ref{mPBE2}) up to order of $g$. (Of course, with the correlation energy given by its first order approximation $\Delta \varepsilon_1(z)$.)  This can be
easily done as follows.  Firstly, let us note that the perturbation solution $\Psi \left( z, z_0(C), C \right)$, whose first order expression was shown in Eq.~(\ref{phi-gr-mn-rg-2}), satisfies Eq.~(\ref{mPBE2}), for arbitrary {\em given} constants $C, \bar{z}_0$.  Note that the same equation would also hold if we replace the parameters $C, z_0(C)$ by functions of $C(z)$
, and $z_0(C(z))$ {\em after the derivatives have been taken}.  Let us further define ``partial derivatives'':
\begin{subequations}
\ba
 \frac{\partial \Psi^R}{\partial z}
&\equiv&
\left. \frac{\partial \Psi}{\partial z}
 (z,C,z_0)
 \right|_{C = C(z),z_0 = z_0(C(z))},
\\
 \frac{\partial^2 \Psi^R}{\partial z^2}
&\equiv&
\left. \frac{\partial^2 \Psi}{\partial z^2}
 (z,C,z_0)
 \right|_{C = C(z),z_0 = z_0(C(z))}. \quad\quad
\ea
\end{subequations}
Our discussion above then shows that
\ba
&-& \!\! \frac{\partial^2 \Psi^R}{\partial z^2}
+\frac{1}{m \! + \! n}
\! \left[
e^{n\Psi^R(z)-n^2 \Delta\varepsilon(z)}
-e^{-m\Psi^R(z)-m^2 \Delta\varepsilon(z)}
\right]
\nonumber\\
&=& O(g^2).
\ea
Therefore the renormalized potential $\Psi^R$ would solve the modified PB if the following identity holds:
\be
\frac{d^2 \Psi^R}{d z^2}
= \frac{\partial^2 \Psi^R}{\partial z^2}
+ O(g^2).
\ee

Now, let us calculate the first order full derivative of the renormalized potential $\Psi^R(z,\bar{z}_0)$ w.r.t. $z$, using Eq.~(\ref{phi-gr-mn-rg-3}) and the chain rule:
\ba
\frac{d \Psi^R}{d z} = \frac{\partial \Psi}{\partial z}
 +  \frac{dC}{dz}
  \left( \frac{\partial \Psi}{\partial C}
 + \frac{d z_0}{d C}
 \frac{\partial \Psi}{\partial z_0}
 \right).
\label{phiR-1st-derivative}
\ea
(Here and below $C$ and $z_0$ are treated as functions of $z$ via Eqs.~(\ref{rz_0}) and (\ref{rc-mn}). )
But the sum inside the bracket in Eq.~(\ref{phiR-1st-derivative}) vanishing is precisely the content of the renormalization group equation Eq.~(\ref{rgeq}).  Hence we have
\be
\frac{d \Psi^R}{d z} = \frac{\partial \Psi}{\partial z} + O(g^2).
\label{d-partial-1st}
\ee
Obviously, if we work out the perturbation series up to infinite order, Eq.~(\ref{d-partial-1st}) would become an exact result, valid up to arbitrary order of $g$.

Let us take one more derivative with respect to $z$:
\ba
\frac{d^2 \Psi^R}{d z^2} &=& \frac{d}{dz}
\left( \frac{\partial \Psi}{\partial z}
 + O(g^2) \right)
 \nonumber\\
&=& \frac{\partial^2 \Psi}{\partial z^2}
+  \frac{dC}{dz} \left( \frac{\partial }{\partial C}
 + \frac{d z_0}{d C}  \frac{\partial }{\partial z_0}  \right)
 \frac{\partial \Psi}{\partial z}
  + O(g^2)
 \nonumber\\
&=& \frac{\partial^2 \Psi}{\partial z^2}
+  \frac{dC}{dz} \frac{\partial }{\partial z}
\left( \frac{\partial \Psi}{\partial C}
 + \frac{d z_0}{d C}  \frac{\partial \Psi }{\partial z_0}
  \right)
  + O(g^2)
\nonumber\\
&=&   \frac{\partial^2 \Psi}{\partial z^2}
+ O(g^2) ,
\ea
where in the third line, we have exchanged the order of partial derivatives and have used the identity:
\be
\frac{\partial}{ \partial z} \frac{d z_0}{d C} = 0.
\ee
This is because $dz_0/dC$ is considered as a function of $C$ and does not explicitly contain $z$.  In the fourth line, we have used again the renormalization group equation Eq.~(\ref{rgeq}).   Thus the renormalized potential indeed satisfies the modified PBE  up to the order of $g$.

\subsection{Renormalized Surface Charge Density and Renormalized Debye Length}
The renormalized potential Eq.~(\ref{phi-gr-mn-rg-3}) contains one undetermined parameter $\bar{z}_0$, which must be fixed by enforcing the boundary condition at $z =0$:
\be
\left. \frac{d \Psi^R}{d z} \right|_{z = 0}
= - \eta.
\ee
Using Eq.~(\ref{d-partial-1st}) and (\ref{phi-gr-mn-rg-2}), we have (with $z$ always set to zero after taking the derivative)
\ba
\frac{d \Psi^R}{d z} &=&
\frac{\partial \Psi^R}{\partial z}
=  \frac{\partial}{\partial z} \Big( \Psi_0
+ g \, \Psi_1  +  g\,C\, \phi_R
\Big)
\\
&=&  \left. \frac{\partial \Psi_0}{\partial z} \right|_0
+ g \left. \frac{\partial \Psi_1}{\partial z}  \right|_0
+ g \, C(z = 0) \, \left. \frac{\partial \phi_R}{\partial z} \right|_0.
\nonumber
\ea
Now the second term vanishes because it is constructed in this way, see Eqs.~(\ref{BC-phi1}) and (\ref{phi-gr-m:n}), whereas the third term vanishes because $C(z)$ does so, see Eq.~(\ref{rc-mn}).  Therefore the physical boundary condition is transformed into the following simple form:
\be
\left.
\frac{\partial \Psi_0}{\partial z}
\right|_{z = 0}
= - \eta.
\ee
Interesting enough, this is identical to the boundary condition Eq.~(\ref{BC-phi0}) we used previously for $\Psi_0(z)$.  Now using the near field asymptotics of $\Psi_0(z)$, Eq.~(\ref{Psi_0-mn}), we find that to the order of $O(g^0)$,
\be
\bar{z}_0 = \frac{2}{n\eta} + O(\eta ^ {-2}).
\label{z_0-bar-eta}
\ee

Let us now analyze the leading order far field asymptotics of the renormalized potential Eq.~(\ref{phi-gr-mn-rg-3}).  The far field asymptotics of the first term can be directly written down using Eq.~(\ref{Psi_0-mn}):
\begin{subequations}
\be
\mbox{1st term} \sim
	c_1^{m,n} \, e^{-\alpha z - \bar{z}_0 },
\label{1st-term}
\ee
To obtain the asymptotics of the second term, we use Eqs.~(\ref{phi-mn-far}) and Eq.~(\ref{S-mn-asympt}):
\be
\mbox{2nd term} \sim
	- \frac{1}{4 \,\alpha} c_1^{m,n} g \, s_{m,n}
	e^{- \alpha z - \bar{z}_0 }.
\label{2nd-term}
\ee
Since we are calculating quantities only up to the order of $g$, we can replace $\alpha$ in the denominator in Eq.~(\ref{2nd-term}) by $unity$ and rewrite the equation as
\be
\mbox{2nd term} \sim
	- \frac{1}{4 } c_1 g \, s_{m,n}
	e^{- \alpha z - \bar{z}_0 }
	+ O(g^2).
\label{2nd-term-2}
\ee

Finally the third term goes asymptotically as
\be
\mbox{3rd term} \sim
g\, c_1^{m,n} \, h_{m,n}(z_0) \,	e^{- \alpha z -\bar{z}_0 },
\label{3rd-term}
\ee
\end{subequations}
where the function $h_{m,n}(z_0)$ is defined in Eq.~(\ref{h_mn-def}).
Note that all three terms Eqs.~(\ref{1st-term}), (\ref{2nd-term}), and (\ref{3rd-term}) are free of secular term and decay with the same length scale $1/\alpha$, which shall be identified with the renormalized Debye length (up to the first order of $g$).

We still need to calculate the function $h_{m,n}(z_0)$ in order to fully determine the far field asymptotics of the renormalized potential.  Since this function appears together with $g$, and since we are only calculating quantities up to the order of $g$, we are allowed to set $g = 0$ inside the definition of $h_{m,n}(z_0)$, Eq.~(\ref{h_mn-def}).  This leads to
\ba
h_{m,n}(\bar{z}_0 ) &=& \!\! \lim_{z \rightarrow \infty}
 \left[
\frac{1}{2}\,  s_{m,n} \,z
+ \!\! \int_{0}^{z} \!\! \phi_L(z', \bar{z}_0 )
S( z' \!+ \! \bar{z}_0 )dz'
\right]
 \nonumber\\
&\equiv&
h^0_{m,n}(\bar{z}_0 ). \label{h_mn-app}
\ea

Note that the large $z$ (IR) divergence in the above integral has already been cancelled by our renormalization procedure.  On the other hand, the integral also exhibits logarithmic divergence as $z_0 \rightarrow 0$ (UV divergence). Using the near field asymptotics of $\phi_L(z)$ and $S(z)$ in the integral, we see that for small $z_0$, it scales as
\ba
&& \int_0^{z_*} \left( - \frac{3n}{8 \pi} \right)
\frac{1}{(z' + \bar{z}_0)^3}
\cdot
\frac{n}{6} (z' + \bar{z}_0)^2 dz'
\nonumber\\
&\sim& \frac{n^2}{16 \pi}
\log \left( \frac{\bar{z}_0}{z_*} \right),
\ea
where $z_*$ is an undetermined small number such that near field asymptotics can be used in the regime $(z_0, z_*)$.  Consequently we expect that the following double limit exist:
\ba
C^{m,n}_s &=& \lim_{\bar{z}_0 \rightarrow 0}
	\lim_{z \rightarrow \infty}
\Bigg[ -\frac{n^2}{16\pi}\log{\bar{z}_0}, \nonumber\\
&+& \int_{0}^{z}
\left( \phi_L(z')S(z')
+  s_{m,n} /2
\right) dz'
\Bigg],
\label{C_m-n}
\ea
and Eq.~(\ref{3rd-term}) can be rewritten into
\be
\mbox{3rd term} \sim
g\, c_1^{m,n} \!\!
\left[
\frac{n^2}{16\pi}\log{\bar{z}_0}
+ C^{m,n}_s
\right]
 e^{- \alpha z - \bar{z}_0 }
 + O(g^2).
\label{3rd-term-2}
\ee

Summing up Eqs.~(\ref{1st-term}), (\ref{2nd-term-2}), and (\ref{3rd-term-2}), we finally obtain the leading order far field asymptotics of the renormalized average potential:
\be
\Psi^R(z) \sim
c_1^{m,n} \!\! \left[ 1
\!+ \! g \left(\frac{n^2}{16\pi}\log{\bar{z}_0}
\! +\! C^{m,n}_s \! -\! \frac{1}{4} s_{m,n} \right)\right]
e^{- \alpha z - \bar{z}_0}.
\label{Psi_R-z}
\ee
The coefficient $\alpha$ therefore is the inverse length scale over which the average electrostatic potential decays in the far field (recall we are using dimensionless units in this work).  It is therefore {\em the ratio between the non-renormalized Debye length and the renormalized one}:
\be
\alpha = \frac{\ell_{\scriptscriptstyle DB } }
{ \ell^{\scriptscriptstyle R}_{\scriptscriptstyle DB}}
= \frac{ \,\, \, \kappa^{\scriptscriptstyle R} } {\kappa}
= 1 +  \frac{1}{2} g \,s_{m,n}  + O(g^2),
\label{m:n alpha-2}
\ee
where $s_{m,n}$ is defined in Eq.~(\ref{s_mn-def}).

In the strongly charged regime, $\bar{z}_0 $ is a small number, and can be neglected in the exponent of Eq.~(\ref{Psi_R-z}). We can further use Eq.~(\ref{z_0-bar-eta}) to express $\bar{z}_0$ inside the logarithm in terms of the bare surface charge density $\eta$, and use Eq.~(\ref{s_mn-def}) to replace $s_{m,n}$.  We finally obtain following result for the {\em one-loop renormalized surface charge density} for a strongly charged plate:
\begin{widetext}
\vspace{-5mm}
\be
\eta_R^{m,n}(\eta,g) =
c_1^{m,n} \left[ 1
- g \left(\frac{n^2}{16\pi}\log{\left( \frac{ n\,\eta} {2}\right)}
- C^{m,n}_s
+ \frac{\log 3}{64 \pi} \,(m-n)^2 \right)\right]
+  O(g^2,\eta^{-1}).
\label{eta_R_mn}
\ee
\end{widetext}
  We can check explicitly that Eq.~(\ref{eta_R_mn}) reduces to Eq.~(\ref{eta_R_11}) for the case $m = n = 1$ (noticing that $c_1^{1,1} = 4$).

Except for the some special cases, we are not able to calculate the constant $C^{m,n}_s$ analytically.  For the cases of $1:-1$, $2:-1$ and $1:-2$, all parts in Eq.~(\ref{C_m-n}) are known explicitly and we can calculate $C^{m,n}_s$ numerically:
\ba
C^{1,1}_s &\approx& 0.005673,\\
C^{2,1}_s &\approx& 0.053428,\\
C^{1,2}_s &\approx& 0.018332.
\ea
Finally let us also quote the corresponding exact results for $c_1^{m,n}$ from \workone:
\ba
c_1^{1,1} &=& 4, \\
c_1^{2,1} &=& 6, \\
c_1^{1,2} &=& 6 (2 - \sqrt{3}).
\ea
The results for case $1:-1$ have of course already been shown in Eqs.~(\ref{eta_R_11}) and (\ref{C_s-11}).

\section{Conclusion and Acknowledgement}
\label{sec:conclusion}

Eqs.~(\ref{eta_R_mn}) and (\ref{m:n alpha-2}) are the main results of this work.  First order renormalization of Debye length by electrostatic correlation in asymmetric electrolytes was studied by Mitchell and Ninham \cite{Mitchell:1968fk} long ago, and our result Eq.~(\ref{m:n alpha-2}) agrees with theirs. In a more recent work, we have also obtained (approximate) analytic result for the renormalized Debye length of the primitive model of asymmetric electrolytes \cite{DLLX-asym-1}, where ions are charged hard spheres, and the density is not necessarily low.  In the limit of low density and zero ion size, this result reduces to Eq.~(\ref{m:n alpha-2}).

\begin{figure}[t!]
	\centering
	\includegraphics[width=8cm]{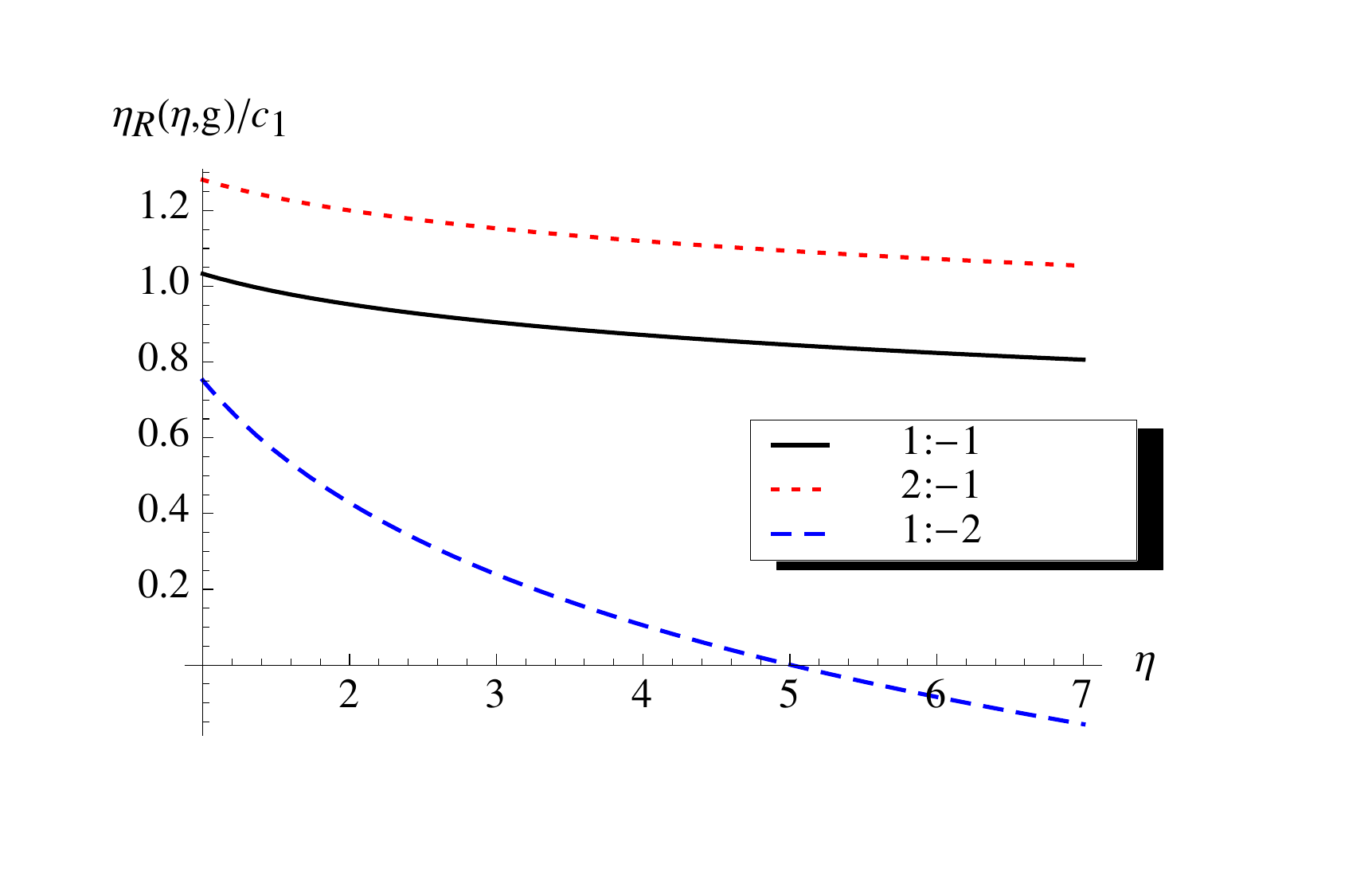}
	\caption{ The renormalized surface charge density as a function of (dimensionless) bare surface charge density $\eta$, Eq.~(\ref{eta_R_mn}), which is defined in Eq.~(\ref{eta-def}).  $b = 7\AA, \ell_{DB} = 15 \AA$.  One can see that divalent counter-ions renormalize the surface charge density substantially downwards and drives charge inversion at $\eta \approx 5$, which corresponds to $\mu \approx 6 \AA$.  }
  \label{fig:plots-eta_R-eta}
\vspace{-3mm}
\end{figure}

\begin{figure}[t!]
	\centering
	\includegraphics[width=8.5cm]{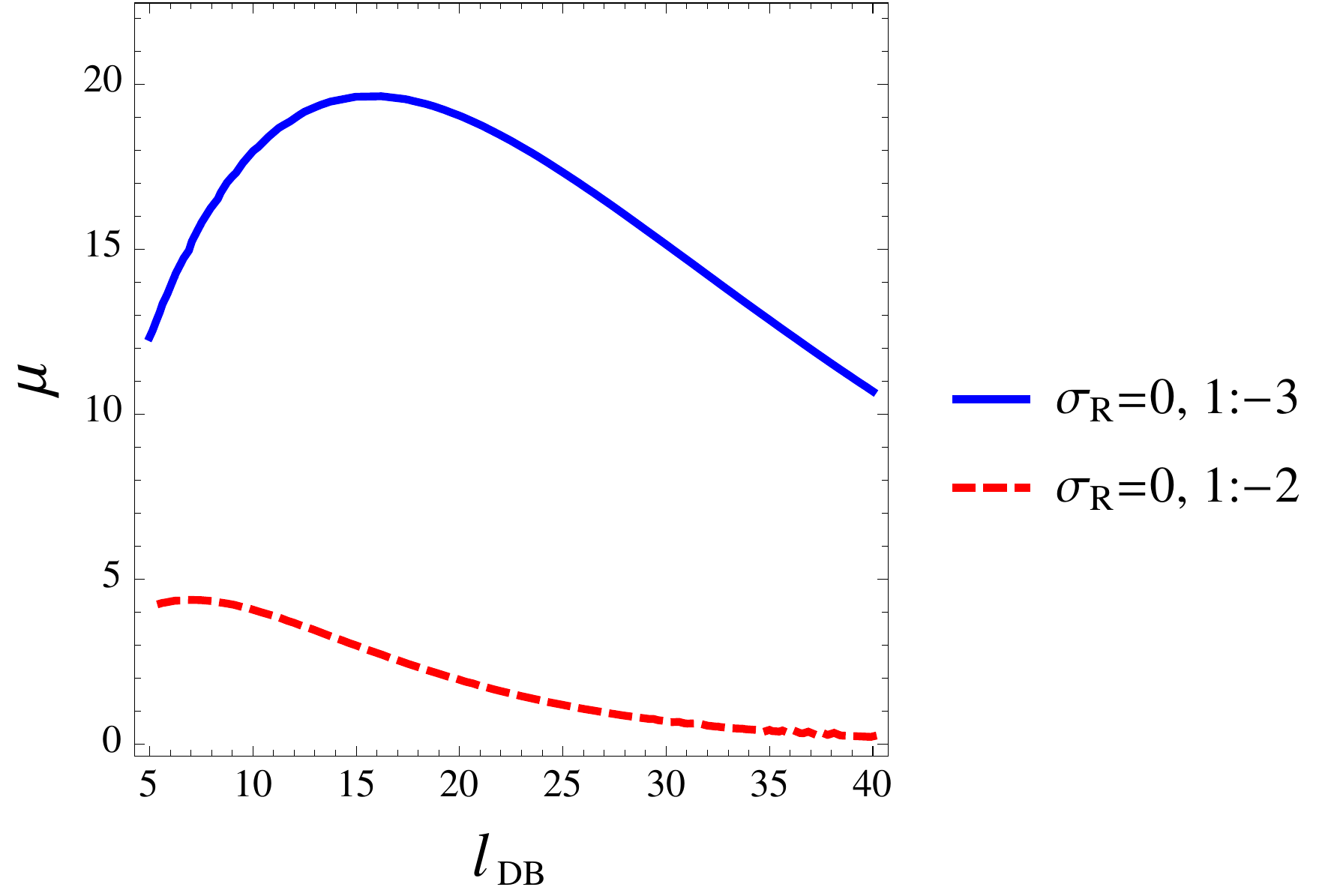}
		\caption{ Locus of vanishing renormalized surface density, according to Eq.~(\ref{eta_R_mn}).
		 $C_1^{1,-3}$ was chosen to be zero as a crude estimate.  Vertical axis: Couy-Chapman length; Horizontal axis: bare Debye length.  Both axes are in the unit of Angstrom.}
  \label{fig:plots-eta_R-2}
\vspace{-3mm}
\end{figure}

Eq.~(\ref{eta_R_mn}) is more interesting because it demonstrates certain general features about the renormalization of surface charge density due to electrostatic correlations:  The leading order renormalization is linear in $g = 4 \pi \ell_{DB}/b$.  In the dilute and strongly charged $g \rightarrow 0, \eta \rightarrow \infty$, and  Eq.~(\ref{eta_R_mn})  reduces to $c^{m,n}_1$, which is the prediction of nonlinear PB theory studied in \workone.  For non-vanishing $g$, the one-loop renormalization contains a negative term logarithmic in $\eta$.  Such a singular term can not be obtained by simple calculations.  Furthermore, the magnitude of this term is proportional to $n^2$, and therefore increases strongly with the valence of counter-ions.  Therefore high valence counter-ions can strongly renormalize the surface charge density downwards, and can drive charge inversion if the bare surface charge density is sufficiently large.  By contrast, the valence of co-ions only appear in the last two terms of  Eq.~(\ref{eta_R_mn}), which are independent of the bare surface charge density.  The valence of co-ions are therefore plays less important role in the renormalization of surface charge density.  In Fig.~\ref{fig:plots-eta_R-eta}, we plot the ratio $\eta_R(\eta, g)/c_1$ Eq.~(\ref{eta_R_mn}) for the case $b = 7 \AA, \ell_{DB} = 15 \AA$, and $\eta$ within the range $(1,7)$.  Note the nonlinear PB theory predicts a flat straight-line $\eta_R(\eta, g)/c_1 \equiv 1$. We can see that in a $1:-2$ (where counter-ions are divalent), the surface charge density are renormalized substantially downwards.  Furthermore, Eq.~(\ref{eta_R_mn}) predicts a charge inversion at approximately $\eta \approx 5$, which corresponds to a Gouy-Chapman length $\mu \approx 6 \AA$.  By contrast, inside a $2:-1$ electrolyte (where the counter-ions are monovalent and the co-ions are divalent ions), the surface charge density is renormalized upwards by correlations.  Finally, in Fig.~\ref{fig:plots-eta_R-2}, we show the locus (in the $\ell_{\rm DB}-\mu$ plane) of vanishing renormalized surface charge densities inside $1:-2$ and $1:-3$ electrolytes.  We use a crude approximation $C_1^{1,3} \approx 0$, since we do not have any better estimate.  It can be seen there that increase of counter-ion valences has pronounced effects in promoting charge inversion.  These results of course agree qualitatively with previous (both experimental and numerical) studies.

We thank NSFC (Grants No. 11174196 and 91130012) for financial support.


\appendix

\begin{widetext}
\section{Details of the Two Speicial Cases: 2:-1 and 1:-2 Asymmetric Electrolytes}
In this appendix we give some results for $2:-1$ and $1:-2$ electrolytes. For these two special case we have closed form of $\Psi_0(z)$, $\phi_L(z)$ and $\phi_R(z)$.

\label{2:-1&1:-2}
For the case of $2:-1$ electrolyte, the solution to PBE is
\ba
\Psi_0(z)&=&
\log\left[
\frac{1+4 \,e^{-(z+z_0)} + e^{-2(z+z_0)}}
{\left( 1-e^{-(z+z_0)} \right)^2}
\right].
\label{phi_0-2:1}
\ea
Expanding to obtain the far field asympytotics according to Eq.~(\ref{Psi_0-mn}), we get the coefficient $c_1^{2,1} = 6$. The full expression for the correlation energy $\delta \varepsilon(z)= g \, \delta\hat{\varepsilon}(z)$ is very complicated.  We shall refer the readers to reference~\cite{xing-bing} for details.  Here we only display its leading order near field and far field asymptotic behaviors
\ba
\delta\hat{\varepsilon}(z)\sim
\left\{
\begin{array}{ll}
{\displaystyle
-\frac{3}{16\pi (z+z_0)},
} \quad
&
z \rightarrow 0;
\vspace{3mm}
\\
{\displaystyle
\frac{3\log{3}}{8\pi} \,e^{- z} ,
}
&
z \rightarrow \infty.
\end{array}
\right.
\label{2:1 ce}
\ea
The function $S(z)$ is related to $\phi_0(z)$ and $\delta \hat{\epsilon}(z)$ via Eq.~(\ref{S_z-def}).
Its far field and near field asymptotics are:
\be
\label{S(z)-2:1}
S(z) \sim \left\{
\begin{array}{ll}
{\displaystyle
-\frac{3}{8 \pi (z+z_0)^3},
} \quad
&
z \rightarrow 0;
\vspace{3mm}
\\
{\displaystyle
-\frac{3\log{3}}{8\pi} \, e^{- z} ,
}
&
z \rightarrow \infty.
\end{array}
\right.
\ee

Two homogeneous solutions to Eq.~(\ref{phih-eqn}) can also be found:
\begin{subequations}
\ba
\phi_R(z)&=& - \phi'_0(z)
= \frac{3\, \coth({(z+z_0)}/{2})}{2 + \cosh(z+z_0)};
\\
\phi_L(z)&=& \frac{1}{12( 2+\cosh(z+z_0))}
\bigg[ 1+10\coth(z+z_0)+\cosh(2(z+z_0)) \nonumber\\
&-& 6 (z+z_0) \coth \left({(z+z_0)}/{2}\right) \bigg]
+ f_1 ( z_0 )  \, \phi_R(z).
\ea
\label{2:1 phiLR}
\end{subequations}
The constant $f_1 (z_0)$ is again determined by the boundary condition Eq.~(\ref{BC-phi1}):
\ba
f_1 (z_0)
&=&
\frac{1}{6}\, z_0 +  \frac{1}{9} \left(1 - \frac{18} {3+2\cosh(z_0)
+ \cosh( 2 z_0 ) } \right)
\sinh(z_0)
+ \frac{1}{36} \sinh(2z_0).
\nonumber\\
 &=& \frac{1}{6} \,z_0^3 + O(z_0^4).
 \label{C_L 2:1}
\ea

For the $1:-2$ electrolyte, the mean potential is
\ba
\Psi_0(z) =
\log \left[ {\frac
{\left(1 + u e^{-(z+z_0)}\right)^2}
{1 - 4u e^{-(z+z_0)} + u^2 e^{-2(z+z_0)} }}
\right],
\ea
where $u = 2- \sqrt{3}$. Hence $c_1^{1,2} = 6 u =  6 (2 - \sqrt{3})$.

The leading order near field and far field asymptotic behaviors of the correlations energy are:
\ba
\delta\hat{\varepsilon}(z)\sim
\left\{
\begin{array}{ll}
{\displaystyle
-\frac{3}{16\pi (z+z_0)},
} \quad
&
z \rightarrow 0;
\vspace{3mm}
\\
{\displaystyle
- \frac{3 u \log{3}}{8\pi} \,e^{- z} ,
}
&
z \rightarrow \infty.
\end{array}
\right.
\label{2:1 ce}
\ea
The function $S(z)$ is related to $\phi_0(z)$ and $\delta \hat{\epsilon}(z)$ via Eq.~(\ref{S_z-def}).
Its far field and near field asymptotics are:
\be
\label{S(z)-2:1}
S(z) \sim \left\{
\begin{array}{ll}
{\displaystyle
-\frac{3}{4 \pi (z+z_0)^3},
} \quad
&
z \rightarrow 0;
\vspace{3mm}
\\
{\displaystyle
-\frac{3u \log{3}}{8\pi} \, e^{- z} ,
}
&
z \rightarrow \infty.
\end{array}
\right.
\ee
The two homogeneous solutions are:
\begin{subequations}
\ba
\phi_R(z) =
\frac{\phi_{Ld}(z)}{12 u
   \left(u+e^{z+z_0}\right) \left(u^2-4 u
   e^{z+z_0}+e^{2
   (z+z_0)}\right)}
 + f_1(z_0)\phi_R(z),
\ea
\end{subequations}
with
\ba
\nonumber
\phi_{Ld}(z) &=& e^{- ( z + z_0)}
   \Big(u^5-9 u^4
   e^{z+z_0}-4 u^3 e^{2
   (z+z_0)} \left(-3 z-3
   z_0+6
   \sqrt{3}+2\right) \\
   &+& 4 u^2
   e^{3 (z+z_0)}
   \left(-3 z-3 z_0+6
   \sqrt{3}-2\right)-9 u e^{4
   (z+z_0)}+e^{5
   (z+z_0)}\Big),
\ea
and
\ba
\nonumber
f_1(z_0) &=&
\frac{1}{72}\Big(
\frac{e^{2 z_0}}{u^2} - \frac{4 e^{z_0}}{u}
+ 4 u e^{-z_0}  - u^2 e^{-2 z_0}
+ 12 \left( -2 \sqrt{3}+z_0 \right)
\Big)
\\
&&
+
\frac{ 2 u e^{z_0} \left(e^{2 z_0}- u^2 \right)
}
{\left( e^{4 z_0} -  2 u e^{3 z_0}  + 6 u^2 e^{2z_0}
- 2 u^3 e^{z_0} + u^4
\right)}.
\ea
\end{widetext}

\end{document}